\documentclass[prd,twocolumn,nofootinbib]{revtex4-1}
\usepackage{amsfonts}
\usepackage{amsmath}
\usepackage{amssymb}
\usepackage{bm}
\usepackage{dcolumn}
\usepackage[dvips]{graphicx}
\usepackage[normalem]{ulem}
\usepackage{latexsym}
\usepackage{rotating}
\usepackage[colorlinks=true]{hyperref}
\usepackage{xspace} 
\usepackage[usenames]{color}
\usepackage[dvipsnames]{xcolor}
\usepackage{mathrsfs}
\usepackage{multirow}
\usepackage{pifont}
\usepackage{enumitem}
\usepackage{color}
\usepackage{url}
\usepackage{appendix}
\usepackage[utf8]{inputenc}
\usepackage{comment}
\usepackage{xcolor}
\usepackage[normalem]{ulem}
\definecolor {darkgreen}{rgb}{0.2,0.7,0.2}
\definecolor{purple}{rgb}{0.5,0,0.5}

\usepackage{graphicx}
\usepackage{epsfig}
\usepackage{slashed}
\usepackage{braket}
\usepackage{tensor}

\usepackage{cases}

\newcommand{\Tr}{{\rm Tr}}

\def\bec{\begin{center}}
\def\eec{\end{center}}
\def\beq{\begin{equation}}
\def\eeq{\end{equation}}
\def\bea{\begin{eqnarray}}
\def\eea{\end{eqnarray}}
\def\muhat{\hat{\mu}}

\def\I{\indices}

\def\MC{\mathcal}
\begin{document}
\title{Tensor network formulation of two-dimensional  gravity }

\author{Muhammad Asaduzzaman}
\affiliation{Department of Physics, Syracuse University, Syracuse, NY 13244, USA.}

\author{Simon Catterall}
\affiliation{Department of Physics, Syracuse University, Syracuse, NY 13244, USA.}

\author{Judah Unmuth-Yockey}
\affiliation{Department of Physics, Syracuse University, Syracuse, NY 13244, USA.}

\begin{abstract}
We show how to formulate a lattice gauge theory whose naive continuum limit corresponds to two-dimensional (Euclidean) quantum gravity including a positive cosmological constant. More precisely the resultant continuum theory corresponds to gravity in a
first-order formalism in which the local frame and spin connection are treated as independent fields. 
Recasting this lattice theory as a tensor network allows us to
study the theory at strong coupling without encountering a sign problem. In two dimensions this
tensor network is exactly soluble and we show that the system has a series of
critical points that occur for pure imaginary coupling
and are associated with first order phase transitions.  We then augment the action with a Yang-Mills term which allows us to
control the lattice spacing and show how to apply the TRG to compute the free energy and look for critical behavior.
Finally we perform an analytic continuation in the gravity coupling in this extended model
and show that its critical behavior in a certain
scaling limit depends only on the topology of the underlying lattice. We also show how
the lattice gauge theory
can be naturally generalized to generate the Polyakov or Liouville action for two
dimensional quantum gravity.
\end{abstract}
\date{\today}
\maketitle

\section{Introduction}

The challenge of formulating a quantum theory of general relativity (GR) remains unmet, and many different approaches have been developed to surmount this challenge. Perhaps the most popular approach gives up on local quantum field theory 
as the appropriate framework and instead posits that the elementary degrees of freedom are strings with GR emerging as the leading low energy effective field theory.
This has been an incredibly productive line of research and has led to a plethora of developments on both physics and mathematics \cite{polyakov103quantum,damour1994string,mukhi2011string, 
smolin2003far, belavin241infinite,mohaupt2003introduction,szabo2002busstepp,polchinski1998string}. Other approaches start from a more conservative
standpoint and take the stance that while quantum gravity is perturbatively non-renormalizable, it may, nevertheless,  be possible to make sense of it
in a non-perturbative context. This is the viewpoint taken by the asymptotic safety program \cite{niedermaier2006asymptotic,reuter1998nonperturbative,lauscher2001ultraviolet,lauscher2002quantum,weinberg1979ultraviolet,weinberg2009living} which employs functional renormalization group methods to search for new fixed points in quantum general relativity.
It is also common to lattice approaches such as causal and Euclidean dynamical triangulations \cite{Ambjorn:2004qm,Ambjorn:2005db,Ambjorn:2005jj,Laiho:2016nlp, Ambjorn:2013eha, Renken:1997na,Catterall:1998gw}.
The latter have revealed  some intriguing aspects of quantum gravity like the emergence of fractal structure at the Planck scale \cite{Ambjorn:2005db,Laiho:2016nlp}.
Loop quantum gravity (LQG) \cite{rovelli2008loop,sahlmann2012loop,rovelli1996loop,rovelli2008loop} which is 
based on Ashtekhar's reformulation of general relativity \cite{sen1982gravity,ashtekar1986new,ashtekar1987new} constitutes
yet another avenue of attack and has led to interesting results like discreteness of spacetime at the Planck scale and a calculation of the entropy of Schwarzschild 
black holes \cite{meissner2004black,rovelli1996loop}.

Some other well-known approaches to quantum gravity are decorated tensor network models \cite{dittrich2016decorated}, spin-foam models of gravity \cite{delcamp2017towards}, twistor theory \cite{penrose1999twistor,penrose1973twistor,penrose1977twistor}, causal sets \cite{bombelli1987space,rideout2001evidence,surya2019causal}, and non-commutative formulations \cite{connes1994noncommutative, seiberg1999string,connes1998noncommutative}.

In this paper we focus on another approach to defining a path integral for
gravity based on its reformulation as a gauge theory, following along the lines 
described in \cite{MacDowell:1977jt,chamseddine1989gauge,smolin1979quantum}. 
Here, we implement this idea in a non-perturbative setting by discretizing the continuum construction as a lattice gauge theory\cite{Catterall:2009nz,caracciolo1988phases}. 

Once one has
a lattice formulation it is natural to attempt to explore the theory using the powerful numerical techniques that have proven so successful in understanding the low energy dynamics of QCD. Unfortunately, as we show later, in the case of gravity the lattice theory typically possesses a sign problem which makes standard Monte Carlo approaches to simulation difficult. However, as we show in this paper, the lattice model can be rewritten as a tensor network and new techniques such as the tensor renormalization group (TRG), which are insensitive to sign problems, can then be brought to bear\cite{levin2007tensor,gu2009tensor,evenbly2015tensor}. There are many different implementations of the TRG algorithm that have been developed but they all have in common the goal of computing the
physical properties of a given network by a recursive procedure that coarsens the tensor network at some scale while attempting to preserve the long distance physics.
The review \cite{evenbly2017algorithms} compares the differences and strengths of these different algorithms. Tensor network formulations have had some success in analyzing many spin and lattice gauge models \cite{Bazavov:2019qih,Liu:2013nsa,PhysRevB.86.045139,Butt:2019uul} but this paper constitutes the first attempt to use them to study quantum gravity.

Implementing any of these TRG schemes is currently infeasible in full four dimensional quantum gravity since the memory required increases rapidly with dimension. Instead, in this paper, we will explore their use in a simpler toy model -- two-dimensional quantum gravity.  Two-dimensional gravity has been well studied using conventional metric-based approaches \cite{Jackiw:1984je,Jackiw:1995hb} including those arising from string theory \cite{Callan:1992rs,Witten:1992fp,Ginsparg:1988ui,Ginsparg:1991bi,seiberg1990notes}. In this paper we try to make contact with these continuum formulations from the discrete tensor network representation. 

We first show that the very simplest transcription of
the continuum theory possesses a tensor network which can be computed exactly {\it without} employing the TRG. Of course to make contact with continuum approaches one needs to
take a continuum limit of the lattice theory. Quite generally this requires that the lattice theory possess a continuous phase transition where the correlation length of the lattice 
fields diverges.
\footnote{Once the 
correlation length expressed in lattice units
becomes large---but smaller than the lattice volume---microscopic details on the scale of the lattice
spacing are washed out and the long distance physics can be captured by
a continuum theory.} One powerful way to
search for such phase transitions is to look for zeroes on the partition function in the plane of complex coupling - so-called
Fisher zeroes. Typically these approach an axis in the thermodynamic limit with the
rate of approach being controlled by a critical exponent. 

We show that in this simplest model there are an infinite number of zeroes but they 
all lie on the axis corresponding to a pure imaginary coupling.
Theories with complex couplings are said to exhibit sign problems and
cannot be studied numerically with Monte Carlo algorithms.
However, TRG algorithms are agnostic to the existence of sign problems since
they do not involve probabilistic sampling and thus can be
used in such cases. We will argue that any gauge theory formulation of gravity 
will possess a sign problem. Thus recasting lattice versions of these
models as tensor networks can be very useful for the numerical study of such theories
at strong coupling.

However, we show that in this simplest model that these transitions are all
first order with a finite correlation length and hence {\it cannot} be used to construct a continuum theory of quantum gravity.

To try to remedy the situation we have considered enlarging the set of operators in the lattice action. Perhaps the simplest such operator which is consistent with all the exact lattice symmetries is the usual Yang-Mills term. Such a term
is not coordinate invariant but is an irrelevant operator which should play no role at long distances. It has the merit however of allowing us to control the lattice spacing 
by tuning its coupling $\beta$. We show how to construct a tensor network representation of this extended theory and use TRG techniques to compute its
free energy and search for phase transitions. Again, a transition is only found for
pure imaginary gravitational coupling.
Furthermore, we are able to recover the results of the TRG analysis using an
exact method based on analytic continuation and find that the critical exponent characterizing
the behavior of the partition function close to the transition depends on the
topology of the lattice. This feature is in qualitative agreement with the Polyakov model of two-dimensional quantum gravity. However, the associated critical
exponent does not agree with the continuum theory. We suggest the discrepancy is the result of neglecting another marginally relevant operator
in the lattice action. We give arguments that the inclusion of such an operator should allow for a precise connection to continuum formulations of two-dimensional quantum gravity. Work in this direction is ongoing \cite{us}.

This paper is organized as follows. In section~\ref{sec2}, we review the first-order formulation of general relativity and discuss its connection to MacDowell-Mansouri gravity in section~\ref{sec3}. We use the ideas developed in these sections to construct a two-dimensional gauge theoretic version of gravity in section~\ref{sec4}. In section~\ref{sec5} we discretize the two-dimensional theory and use the character expansion of the associated class 
function to find an analytic expression for the partition function. Using the closed-form results obtained in the previous section, we perform a Fisher zero analysis to determine the order of the phase transition in Sec.~\ref{sec6}. In section~\ref{sec7}, we argue that we can add an irrelevant Yang-Mills operator to the lattice action to control the lattice spacing. This augmented model
is then explored using the TRG. In section~\ref{TRG} we perform an analytic continuation in the gravity coupling and
are able to show that the critical exponents characterizing the critical behavior
depend only on the topology of the lattice.
We argue in section~\ref{POLY} that a connection to the continuum Polyakov
theory of two-dimensional gravity can be made only after the inclusion of a new marginal operator and we describe that connection in some detail.
Finally, we conclude our discussion in Sec.~\ref{sec8} with possible extensions of the work by
writing down a lattice theory of 4D quantum gravity that parallels the two-dimensional theory in this
paper.

\section{\label{sec2}Review of the Palatini-Cartan formulation of Einstein gravity}
A path integral for Einstein-Hilbert gravity written in terms of a metric $g_{\mu\nu}$
takes the form
\beq
Z=\int Dg e^{iS_G(g)}\eeq
where the action $S_G$ is given by
\beq
S_G = \frac{1}{\ell_{P}^{2}} \int d^4x\,\sqrt{-g}\left(R-2\Lambda\right)\eeq
It is known how to map gauge theories into tensor networks by following the procedure described
in eg. \cite{Bazavov:2019qih}. This method, based on character expansion of the
group valued action, will be described in detail in section~\ref{sec4}. However,
there is no analogous method known for non-compact spin two
fields such as $g_{\mu\nu}$. Fortunately, gravity can be recast in the
language of Yang-Mills theory by replacing the metric $g_{\mu\nu}(x)$ by the so-called frame field $e_\mu(x)$  by employing the relation
\beq
g_{\mu\nu}=e\I{_\mu^a} e\I{_\nu^b} \eta_{ab}\label{metric}\eeq
\cite{eguchi1980gravitation, Yepez:2011bw}.
Clearly the metric is invariant under {\rm local} Lorentz rotations of the frame field $e_\mu^a(x)\to  \Lambda^a_b(x)e_\mu^b(x)$.
In order to write down derivative terms one must also
introduce a corresponding gauge field corresponding to these
local Lorentz transformations---the spin connection 
$\omega_\mu(x)$ transforming as
\beq
\omega\I{_\mu^{ab}}\to D\I{_\mu^{ac}} \phi^{cb}=\partial_\mu\phi^{ab}+\left[\omega_\mu,\phi\right]^{ab}\eeq
where $\omega_\mu=\sum_{a<b}\omega_\mu^{ab}(x)T^{ab}$ is summed over the generators of the Lorentz
group.
A natural locally Lorentz invariant action can then be written down in terms of $e_\mu$ and the usual Yang-Mills
curvature $R_{\mu\nu}=[D_\mu,D_\nu]$ as
\begin{align}
S=\frac{1}{\ell_P^2}\int d^4x \, \epsilon^{\mu\nu\rho\lambda}\epsilon_{abcd}\big (e\I{_\mu^a} e\I{_\nu^b}  R\I{_{\lambda\rho}^{cd}} 
-\frac{1}{\ell^2} e\I{_\mu^a} e\I{_\nu^b} e\I{_\lambda^c} e\I{_\rho^d} \big)\label{palatini}
\end{align}

Notice that since $g_{\mu\nu}$ is not a fundamental field in this approach, the only tensor available to contract world indices is the invariant tensor $\epsilon_{\mu\nu\rho\lambda}$ which automatically guarantees
that the theory is independent of coordinate transformations. Rather remarkably this action reduces to the
usual Einstein-Hilbert action provided
\begin{align}
{\rm det} (e\I{_\mu^a})\ne 0\\
T\I{_{\mu\nu}^a}=D_{\left[\mu\right.}e\I{_{\left.\nu\right]}^a}=0 \label{torsion}
\end{align}
where the first line guarantees that we can invert the frame field considered as a $4\times 4$ matrix and the second is the usual vanishing torsion condition required to achieve a theory that depends only on the metric by supplying an additional condition that expresses the spin connection in terms of the frame field.  Eq.~\eqref{torsion} is a classical equation of motion of the action in Eq.~\eqref{palatini}. The first term in the Eq.~\eqref{palatini} reduces to the usual Ricci scalar of the metric theory once one employs the relation $e\I{^\lambda_a}e\I{_\lambda^b}=\delta\I{^b_a}$, and the result 
$\epsilon^{\mu\nu\rho\lambda}={\rm det}\,(e)\epsilon^{abcd}e^\mu_a e^\nu_b e^\rho_c e^\lambda_d$,
\beq
\epsilon^{\mu\nu\rho\lambda} \epsilon_{abcd} e\I{_\mu^a} e\I{_\nu^b} R\I{_{\lambda\rho}^{cd}}\to \sqrt{-g} \, R. 
\eeq
while the second is clearly the cosmological constant term ${\rm det}(e)=\sqrt{{\rm det}(g)}$
with $\Lambda=\frac{1}{2 \ell^2}$ the cosmological constant. Notice the introduction of an arbitrary length scale $\ell$ to keep the frame field dimensionless.

This Cartan-Palatini formulation
of gravity offers several advantages over the conventional metric approach: it is explicitly
independent of any background metric, it employs the familiar formalism of
gauge theories and it can naturally include fermions using the spin connection. Its main disadvantage is that it contains two independent fields---the frame and spin connection. However we will
see in the next section that it is possible to enlarge the gauge symmetry in such a way that both fields play an equivalent role.

\section{\label{sec3}Cartan gravity as spontaneously broken de Sitter gravity}
While this gauge theoretic form of gravity is clearly an improvement over the metric
based formulation from the point of view of constructing a tensor network it still contains
a non compact field $e_\mu$ in addition to the gauge field $\omega_\mu$ which prevents a
simple application of the method of character expansion to build
the tensor network. Luckily, there
is a way to enlarge the gauge symmetry and treat $e_\mu$ as another gauge field.
Once we go to the lattice
this will allow the standard character expansion approach to be employed.

The procedure was first described in the continuum in
\cite{MacDowell:1977jt} and requires extending
the Lorentz group to the de Sitter group with $e_\mu$ playing the role
of the additional gauge fields. As we will see the key physical requirement that allows this
new theory to reduce to Einstein-Hilbert is that the vacuum state of the theory 
must correspond to a phase in which the de Sitter symmetry is 
spontaneously broken down to Lorentz symmetry.

For simplicity we will restrict the discussion from this point to the Euclidean theory in which the $SO(4)$ Lorentz symmetry is embedded in the larger $SO(5)$ Euclidean de Sitter symmetry. The action that is required is
\begin{align}
S_M=\kappa \int d^4x\, \epsilon^{\mu\nu\lambda\rho} \epsilon_{ABCDE}\, \phi^E F_{\mu\nu}^{AB}
F_{\lambda\rho}^{CD} \label{MM}
\end{align}
with $ A,B,\ldots=1\ldots 5$. The curvature $F$ is the usual $SO(5)$ Yang-Mills term while the scalar field is a new degree of freedom which transforms in the fundamental representation of $SO(5)$. We can use the local $SO(5)$ transformations to set four of its five
fields equal to zero and write $\phi^A=\rho(x)\delta^{5A}$. This is called
working in unitary gauge and does not change the gauge invariant physical content of the
theory.
We can decompose this action under the remaining
$SO(4)$ subgroup that is preserved by this gauge choice. This subgroup
can then be identified with the Lorentz symmetry and the corresponding gauge field with the spin connection $\omega_\mu$. The four gauge fields corresponding to the additional generators of $SO(5)$ correspond to the frame field $e_\mu$
\beq
 A_\mu=\omega_\mu^{ab}T^{ab}+\frac{1}{\ell}e_\mu^{a}T^{5a}
\eeq
with  $a,b=1\ldots 4$.
In a similar way the $SO(5)$ curvature decomposes under the $SO(4)$ subgroup  as
\beq
F_{\mu\nu}=(R_{\mu\nu}^{ab}-\frac{1}{\ell^2}e_{\left[\mu\right.}^a e_{\left. \nu\right]}^b)\,T^{ab}+D_{\left[\mu\right.} e^a_{\left.\nu\right]}\,T^{5a} \eeq
where $R$ is the $SO(4)$ curvature as before. Finally, we assume 
a non-zero constant expectation value for the radial
scalar field $\langle \rho(x) \rangle = \rho_0$ in the vacuum state. A vacuum expectation value of this form
is only invariant under the generators corresponding to the Lorentz subgroup and hence de Sitter 
symmetry is spontaneously broken down to Lorentz symmetry.
With this additional condition, it is straightforward to see
that the $SO(5)$ action given in Eq.~\eqref{MM} reproduces the action of GR in Eq.~\eqref{palatini} provided
$\kappa=\left(\frac{\ell}{\ell_P}\right)^2\sim \frac{1}{G\Lambda}$\footnote{An additional topological term - the Euler
density is also produced which we ignore since we are only concerned with manifolds of fixed Euler
character}.

There are several advantages to this construction. Firstly, the classical equations of motion of the $SO(5)$ theory
ensure that both $F_{\mu\nu}=0$ and $D_\mu\phi=0$. The former implies that both the Einstein equation and
the torsion free condition are satisfied classically
\begin{align}
R_{\mu\nu}^{ab}-\frac{1}{\ell^2}e_{\left[\mu\right.}^a e_{\left. \nu\right]}^b&=0,\\
D_{\left[\mu\right.} e^a_{\left.\nu\right]}&=0.\end{align}
Notice that the vanishing of the covariant derivative of the scalar is also consistent
with the condition for spontaneous symmetry breaking. 

The enlarged gauge symmetry also helps to constrain counter terms in the theory and requires
that the correct measure for path integration be invariant under the de Sitter symmetry\footnote{It should be noted
that essentially the same construction works in odd dimensions and generates Witten's representation of 3d gravity as a Chern-Simons gauge theory. In that case however there is no need for an additional scalar to break the symmetry \cite{Witten:1988hc}}.

\section{\label{sec4}Two dimensions}
Having reviewed how this procedure works in four dimensions we now
devote the remainder of this paper to an exploration of the approach in the case of two dimensions. There are several studies of gauge theoretic formulations of gravity in two dimensions \cite{fukuyama1985gauge, isler1989gauge, chamseddine1989gauge, chamseddine1990topological} and a great deal is known about
the metric theory through Liouville theory and matrix models \cite{Ginsparg:1988ui,Ginsparg:1991bi}.  In two dimensions,
the Einstein-Hilbert action is a topological invariant\footnote{$\int \sqrt{g} R=4\pi\chi$
with $\chi$ the Euler character of the manifold 
which counts the number of handles of the manifold and is independent of $g$.
Topological theories possess only a finite number of degrees of freedom in contrast
to a generic quantum field theory} and the Einstein tensor is identically zero. To obtain a non-trivial analog of the Einstein equation, Jackiw and Teitelboim proposed a solution, $R-2 \Lambda=0$, where $R$ is the Ricci scalar and $\Lambda$ is the cosmological constant.  The proposed Lagrangian of Jackiw-Teitelboim (dilaton) gravity \cite{Jackiw:1992bw} is,
\begin{align}
\MC{L}=\sqrt{-g} \, \phi \,  (R -\Lambda). \label{eq1} 
\end{align}
Note the introduction of the scalar field which acts as a Lagrange multiplier needed to enforce the equation of
motion. It is analogous to the scalar that appeared in the previous gauge theoretic approach to gravity. 

Returning to this latter construction it is easy to see that the analogous Lagrangian in two
dimensions is,
\beq
S=\int d^2x\, \epsilon^{\mu\nu}\epsilon_{abc} \phi^c F_{\mu\nu}^{ab},\quad a,b=0,1,2\label{2dact}\eeq
where $F$ takes its values in $SO(3)$. Exploiting the homomorphism $SO(3)\sim SU(2)$ this can be rewritten as
\beq
S=\int d^2x\, \epsilon_{\mu\nu} \Tr(\phi F_{\mu\nu})\eeq where $\phi$ is now in the adjoint of $SU(2)$. 
Picking a unitary gauge again allows us to simplify the action further to
\beq
S=\kappa\int d^2x\, \epsilon_{\mu\nu}\rho(x) \Tr(\sigma_3F_{\mu\nu})\label{unitary}\eeq
which clearly exhibits the remaining exact $SO(2)\sim U(1)$ Lorentz symmetry corresponding to transformations
$F_{\mu\nu}\to e^{i\alpha(x)\sigma_3}F_{\mu\nu}e^{-i\alpha(x)\sigma_3}$.  Again the classical equations correspond to vanishing torsion and $R=\frac{2}{\ell^2}$.
Thus we have shown that a natural candidate for two-dimensional gravity takes the form of an
$SU(2)$ gauge theory.  It is then natural to discretize it on a lattice and 
study it numerically at strong coupling. We turn to this in the next section.

\section{\label{sec5}Lattice model and tensor network representation}
The theory can be discretized on any lattice and as a starting point we pick a simple square lattice, placing group
elements $U_\mu(x),\mu=1,2$ of $SU(2)$ on the link $x\to x+\hat{\mu}$ as in lattice QCD \cite{PhysRevD.10.2445}. 
The lattice action takes the form
\beq
S = - \frac{\kappa}{2} \sum_x \Tr (i\phi(x) \left[U_{12}(x)-U_{12}(x)^\dagger\right])\eeq
where $U_{12}(x)$ is the usual Wilson plaquette operator corresponding to the
product of link fields around a unit square
of the lattice with corner at lattice site $x$,
\beq
U_{12}(x)=U_1(x)U_2(x+\hat{1})U^\dagger_1(x+\hat{2})U^\dagger_2(x),
\eeq
where $\hat{1}$ represents a unit vector in the 1-direction etc. Formally expanding
the link fields in powers of the lattice spacing $U_\mu=e^{iA_\mu a}=I+iA_\mu a+\ldots$
reproduces the continuum action Eq.~\eqref{2dact}.

It is convenient to adopt a generalized polar representation of the
adjoint field $\phi(x)=\frac{1}{\sqrt{2}} \rho(x) G(x)\sigma_{3} G(x)^\dagger$ where
$G(x)$ is a general position dependent $SU(2)$ matrix. 
One can then use cyclic invariance of
the trace to rotate the factors of $G$ onto the original link fields and write the
theory in terms of new link fields $U^\prime_\mu(x)=G^\dagger(x)U_\mu(x)G(x+\mu)$. This procedure explicitly effects unitary gauge for the lattice theory.
Finally, dropping the prime labels allows us to write the action as
\beq
S = -\frac{\kappa}{\sqrt{2}}\sum_x \rho(x){\rm Re}\Tr (i\sigma_3U_{12}) \label{2dLA} \eeq
So far we have not changed the physical content
of the theory. To target the theory with just a residual $U(1)$ Lorentz invariance
requires, as before, 
the additional assumption that $\rho$ picks up a vacuum expectation value (VEV). This is done by hand by
setting $\rho(x)=\sqrt{2}$. Since the VEV of $\phi$ lies in the $\sigma_3$ direction
it should  be clear that the original $SU(2)$ is now broken to $U(1)$ corresponding to
rotations of the form $e^{i\alpha\sigma_3}$ with $\sigma_3=i\sigma_1\sigma_2$ corresponding
to rotations in the $12$-plane.

Now, the partition function of the action in Eq.~\eqref{2dLA} is a class (group
invariant) function and by a fundamental theorem of group
theory can be expanded as a sum
over characters with each character corresponding
to the trace of the element of the group in a particular
representation \cite{drouffe1983strong}. Ordinary Fourier transformation
can be envisaged as the analog of this procedure for the $U(1)$ group
where the familiar sines and cosines 
represent all the one dimensional characters of the $U(1)$ group.

To construct the tensor network we employ a similar character expansion
for the $SU(2)$ group and expand the Boltzmann
weight $e^{-S}$ corresponding to each plaquette term in the action
\beq
e^{\kappa\Tr({ MU_{12}})}=\sum_{j = 0}^{\infty} \frac{2(2j+1)I_{2j+1}(\kappa)}{\kappa}
\chi^j\left(MU_{12}\right)\label{char}\eeq
with $M=i\sigma_3=e^{i\frac{\pi}{2}\sigma_3}$ a constant $SU(2)$ matrix and $I_n$ a modified Bessel function is the expansion coefficient. 
The sum runs over all irreducible representations of $SU(2)$ labeled by $j$. 
Expanding the character $\chi^j$
on products of Wigner D-matrices yields an expression for the partition function:
\begin{widetext}
\beq
\label{eq:grav-expand}
Z=\int \prod_l dU_l\,\prod_p \sum_{j} \frac{2(2j+1)I_{2j+1}(\kappa)}{\kappa}D_{ab}^j(M)D_{bc}^j(U_1)D_{cd}^j(U_2)D_{de}^j
(U_3^\dagger)D_{ea}^j(U_4^\dagger)\eeq 
\end{widetext}
with $U_1$, $U_2$ etc \ldots denoting the links around a given plaquette, $\prod_{l}$ a product over links, and $\prod_{p}$ a product over plaquettes. For a two-dimensional torus we can then integrate out the individual gauge links $U_\mu(x)$ using the
result $\int dU D^j_{ab}D^{*k}_{cd}=\frac{1}{2j+1}\delta_{jk}\delta_{ac}\delta_{bd}$. Clearly the result
of this integration ensures that only a single representation survives over the
entire lattice and the resulting expression can be
organized as a product over all sites, $s$,
\beq
Z=\sum_j \prod_{s=1}^{N_{0}} \left[\frac{2I_{2j+1}(\kappa)}{(2j+1)\kappa}\right]D_{aa}^j(M).\eeq
Using the well-known formula for the character of an $SU(2)$ representation \cite{drouffe1983strong} 
\beq
\chi^j=\frac{\sin{\left((2j+1)\theta \right)}}{\sin{\left(\theta \right)}}.
\eeq
with $\theta=\frac{\pi}{2}$ yields
\beq
\label{eq:pure-z}
Z = \sum_j f_{j}^{N_{0}}
\eeq
where
\beq
f_j=\frac{1}{2j+1}\left(-1\right)^j\frac{2I_{2j+1}\left(\kappa\right)}{\kappa},\quad {\rm for\;} j\;{\rm integer}.\eeq  With the partition function written in this form, an obvious tensor network can be built.  Consider the tensor located at lattice plaquette $x$,
\begin{equation}
	T_{i j k l}^{(x)} = \begin{cases}
		f_r & \text{ if $i = j = k = l = 2r$} \\
		0 & \text{ otherwise.}
	\end{cases}
\end{equation}
where each index is shared between two adjacent plaquettes, of which there are four surrounding a single plaquette in two-dimensions.
This tensor is diagonal, with the only nonzero entries being those where all four indices are identical.  By contracting this tensor with itself one reconstructs the above
partition function,
\begin{align}
	Z &= \bigg(\prod_{n=1}^{2N_{0}} \sum_{i_n = 0}^{\infty}\bigg) T_{i_1 i_2 i_3 i_4} \cdots T_{i_{2N_{0}-3} i_{2N_{0}-2} i_{2N_{0}-1} i_{2N_{0}}} \nonumber\\ &= \Tr\bigg[\prod_{x=1}^{N_{0}} T^{(x)}\bigg]
\end{align}
with the trace being interpreted as a tensor trace.  Since the tensor is diagonal, the $N_{0}$ tensors simply reproduce the $N_{0}$th power of the $f_j$s, and the $2N_{0}$ sums for each link, simply reduce to a single sum over representations.  

Critical points of the system correspond to zeros of $Z$ in the plane of complex coupling $\kappa$. In general phase
transitions occur when these so-called Fisher zeros pinch the axes in the thermodynamic limit \cite{Meurice:2012sj,PhysRevD.85.056010}. In the next section
we examine this in more detail.

\section{\label{sec6}Fisher zeros}
\begin{figure}[tbp]
\centering
\includegraphics[scale=0.65]{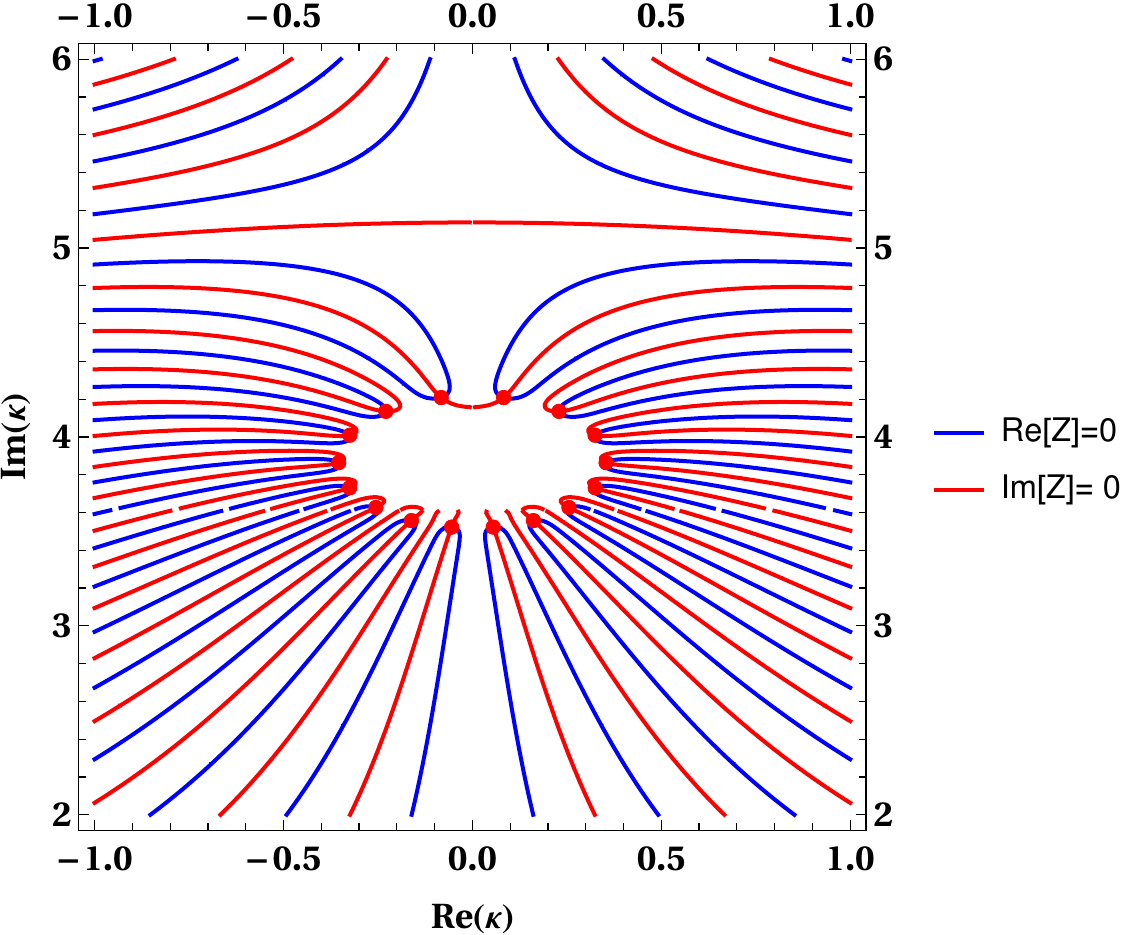} 
\caption{Zeros of the partition function in the complex $\kappa$ plane with $j_{max}=2$, for $N_{0}=16$.}
\label{FZ16}
\end{figure}

\begin{figure}[tbp]
\centering
\includegraphics[scale=0.65]{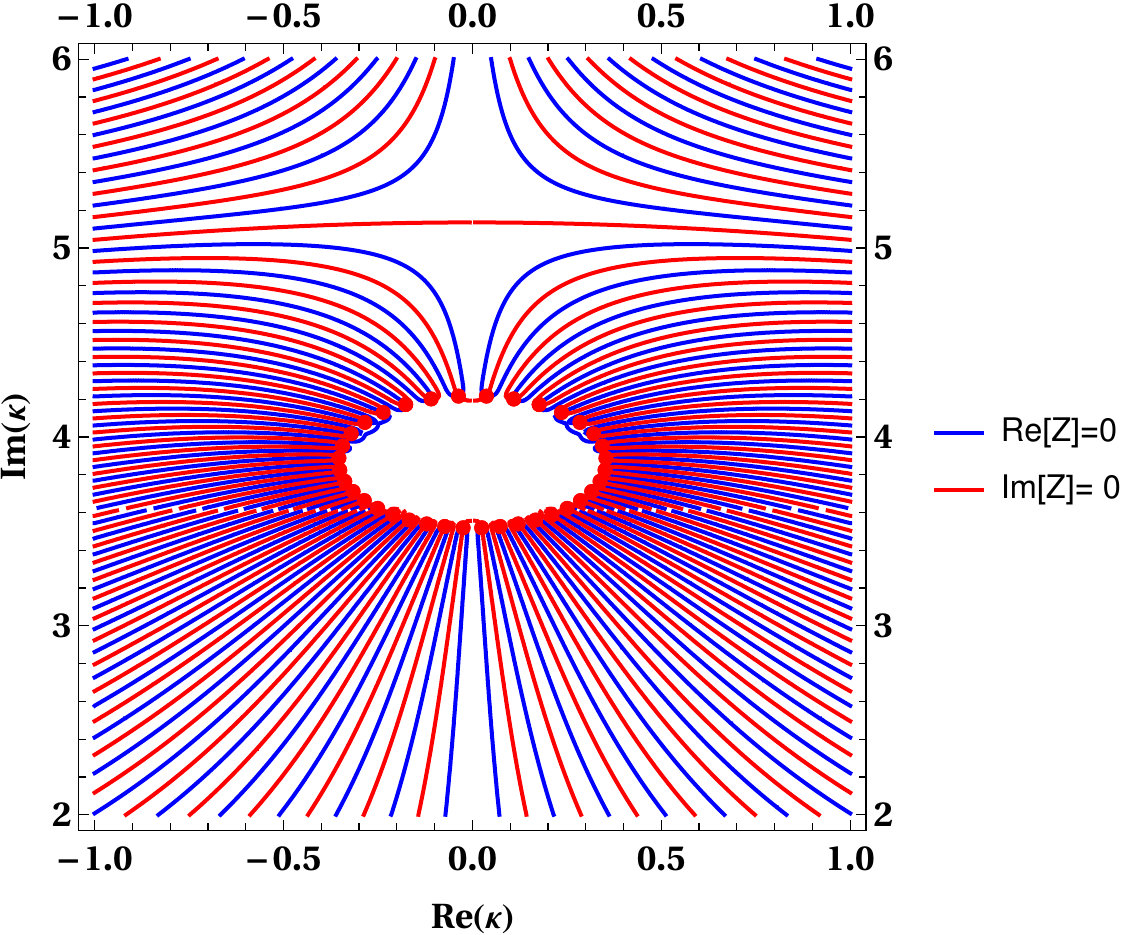} 
\caption{Zeros of the partition function in the complex $\kappa$ plane with $j_{max}=2$, for $N_{0}=36$.}
\label{FZ36}
\end{figure}

In this section, we use the representation from Eq.~\eqref{eq:pure-z}, and with $\kappa$ allowed to take all values in the complex plane we search for zeros of the partition function numerically.

In practice we truncate the expansion in representation $j$ at some $j_{max}$. Setting $j_{max}=2$ we show in Fig.~\ref{FZ16} lines where the real and imaginary parts of $Z$ vanish when $N_{0}=16$.\footnote{Since the successive terms in the expansion drop rapidly with $j$
the truncation to $j_{max}=2$ 
is actually quite good and we have checked that including higher order
terms does not change the conclusions.} Where these curves cross
corresponds to zeros of $Z$. We observe that rings of zeroes develop centered at discrete intervals along the imaginary
$\kappa$ axis. If we focus on the leading ring we can see that the density of zeroes along the ring increases
with $N_{0}$---see Fig.~\ref{FZ36} which shows results for $N_{0}=36$. Indeed, we observe that the number of zeroes is precisely $N_{0}$. Notice that while there are
no zeroes on the imaginary axis the set of zeroes approach the imaginary axis as $N_{0}$ increases. This is precisely
the behavior required of a Fisher zero in the thermodynamic limit. The only twist is that
usually the zeroes approach the real axis while here 
the zeroes approach the imaginary axis implying that the system only develops a phase
transition when the coupling $\kappa$ is pure imaginary. On reflection this actually should not be surprising; the
gravity term resembles a topological term since it employs an epsilon tensor to contract spacetime indices. On Wick rotation to Euclidean space such a term naturally acquires a factor of the square root of minus one.

Notice that such an action with pure imaginary coupling
would be {\it impossible} to simulate using Monte Carlo methods because of a dramatic
sign problem highlighting the advantages of the tensor network approach we adopt here. 

\begin{figure}[tbp]
\centering
\includegraphics[scale=0.7]{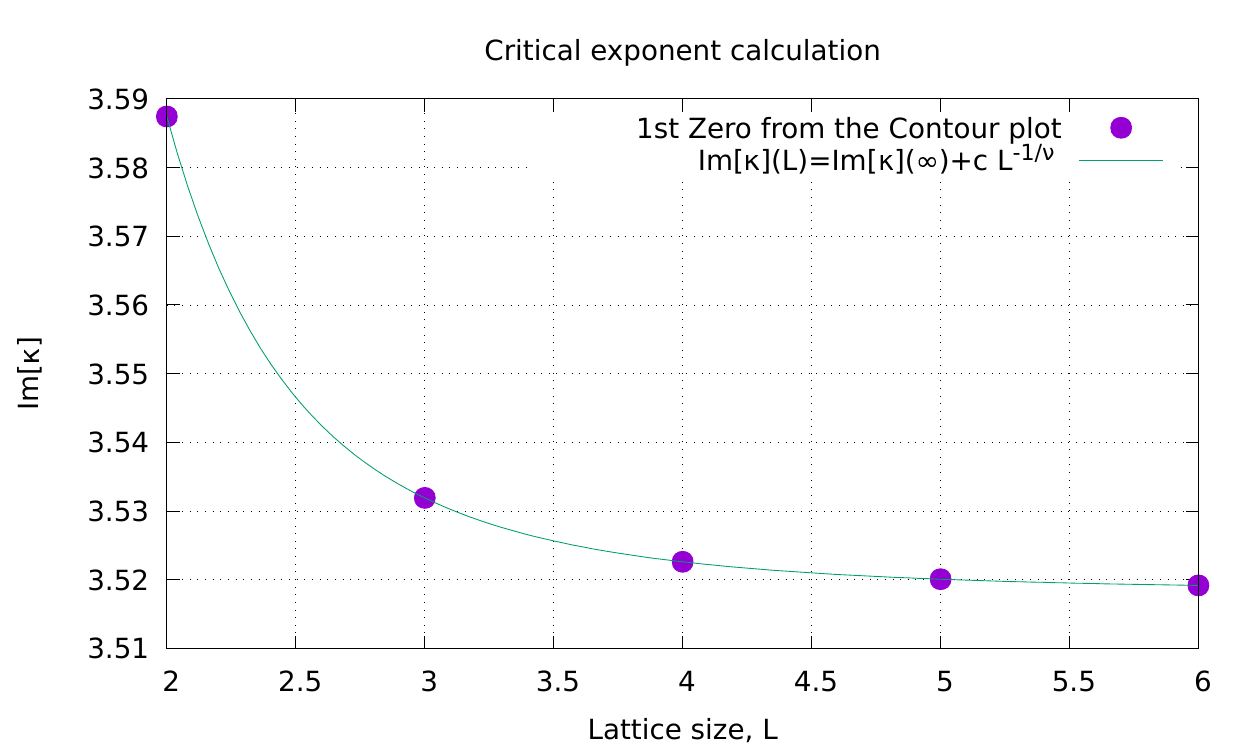} 
\caption{\label{scale} Correlation length exponent ($\nu$) calculation from the zeros of the contour plots in the $\Re[\kappa] \to 0$ limit.}
\end{figure}
Actually the rate at which the zeroes approach the imaginary axis yields the correlation length exponent $\nu$ associated with the phase transition that arises in the thermodynamic limit \cite{Klaus:1997be},
\beq
\kappa_{\rm zero}(N_{0})=\kappa_c(\infty)+A N_{0}^{-\frac{1}{2\nu}}\label{nu}.\eeq
Figure~\ref{scale} shows a plot of ${\rm Im}\,\kappa$ for the zero closest to the axis as a function of
$L=\sqrt{N_{0}}$ together with a fit to the form given in Eq.~\eqref{nu}. The fitted exponent is $\nu=0.2495$ and $\kappa_c(\infty)=3.51833$.

Let us now try to understand this structure using analytical arguments.
In the limit $N_{0}\to\infty$ the
partition function formally truncates to just the leading term $j_{max}=0$. The free energy is then
\beq
f=\frac{1}{N_{0}}\ln{Z}=\ln{\frac{I_1(\kappa)}{\kappa}}.\eeq
This clearly possesses no zeroes on the real $\kappa$ line. However,
taking $\kappa\to i\kappa$ takes $I_1(\kappa)\to J_1(\kappa)$
and it appears that
the free energy possesses a series of logarithmic singularities along the imaginary axis corresponding to the
zeroes of the first Bessel function.
Furthermore
\beq
\frac{\partial f}{\partial\kappa} = \langle {\int \rm det} \,e \rangle = \frac{1}{J_1(\kappa)}\left(J_0(\kappa)-J_2(\kappa)\right)-\frac{1}{\kappa}\eeq
where we have neglected the Riemann term as it is a topological invariant and does not scale with $N_{0}$. Clearly
at points where $J_1=0$ the mean area measured in units of the lattice spacing diverges. 

However while the theory truncates to the leading term $J_1(\kappa)$ for generic values of $\kappa$ in the large $N_{0}$ limit this procedure fails precisely in regions close to the zeroes of $J_1$. 
If the second term $J_3$ is kept in the analysis it is easy to see that 
\begin{align}
Z&=\left(\frac{2}{\kappa}J_1(\kappa)\right)^{N_{0}}, \quad \mathrm{for} \quad \left|\frac{3J_1}{J_3}\right|>1 \\
Z&=\left(\frac{2}{3\kappa} J_3(\kappa)\right)^{N_{0}}, \quad \mathrm{for} \quad \left|\frac{3J_1}{J_3}\right|<1.
\end{align}
Since the zeroes of $J_3$ never coincide with those of $J_1$ there will always be windows in $\kappa$ around
each zero of $J_1$ where the free energy changes from behaving like $J_1$ to $J_3$ and an exact
zero is avoided. The upper and lower
limits of this window can be found by solving the equation $\left|\frac{J1}{3J_3}\right|=1$. Since the Bessel functions
are analytic this interval becomes a curve in the complex plane corresponding to the rings
observed in the Fisher zero analysis.  The occurrence of $N_{0}$ zeroes then corresponds to the solutions
of $\left(3J_1(z-z_0)/J_3(z-z_0)\right)^{N_{0}}=1$  with $z_0$ a zero of $J_1$. Close to $z_0$ this can be approximated
by a linear function of $z-z_0$ which is then proportional to an $N_{0}^{\rm th}$ root of unity. It is interesting to note that from this expression we obtain a critical coupling constant $\kappa_c=3.51832$ where the phase transition occurs. This matches exactly from the finite size scaling analysis described in the first part of this section.  

It should be clear that while the free energy is continuous at this boundary in the large $N_{0}$ limit its
derivative will not be---the jump in the slope being $\frac{\partial}{\partial \kappa} \ln(J_3(\kappa)/J_1(\kappa)) $. Thus one expects a series of finite jumps in the value of $\langle {\rm det\,}e \rangle$ as a function of (imaginary) coupling
$\kappa$. The existence of such first-order phase transitions hence preclude the existence of a continuum
limit in this lattice theory. This conclusion remains even for larger $j_{max}$. Inclusion of the higher order Bessels does not change the contour plots of the zeros of the partition function significantly near the first zero of $J_1$. The largest terms in the expansion
near the first zero of $J_1$ arise from $J_3$ and all others are exponentially suppressed as $N_{0}\to\infty$. It is
possible that some higher Bessels $J_m$ and $J_n$ with $m,n> 3 $ will dominate near some zero of $J_1$ further from the origin so
that the window of convergence will be controlled by $(n\,J_m(\kappa))/(m\, J_n(\kappa))$. However the essential conclusion of
a discontinuous first derivative of the free energy there will continue to hold.

The overall conclusion is that the lattice model based on the classical two-dimensional gravitational action possesses only first order phase transitions.  The absence
of a diverging correlation length ensures that the lattice model
does not possess a continuum limit. Hence, although we started with a naive discretization
of a continuum action the result of including quantum fluctuations has led to a theory
in which it is not possible to take the continuum limit. In the continuum the inability
to remove a UV cutoff is a sign of non-renormalizability of the theory. Often this
is a signal that we have left out of our bare action important operators whose couplings must
be tuned to remove the cutoff---here taking the lattice spacing to zero or
equivalently the correlation length in lattice units to infinity.

The question that then
arises is whether it is possible
to augment the action by additional terms to generate a continuous phase transition and hence the possibility of
a continuum limit. In the continuum there are no additional
relevant or marginal operators that preserve
both the gauge symmetry and coordinate invariance.\footnote{Remember that in this
formulation the metric is emergent and generated by the frame fields and so 
coordinate invariant terms must be built using only the epsilon tensor.}
However, on the lattice coordinate invariance is lost and additional operators
can appear.  Perhaps the simplest of these is a Yang-Mills term ${\rm Tr}\,(F_{12}F_{12})$. Such a Yang-Mills term, while naively
unimportant in the infrared limit that determines the continuum theory,
can play a role in the ultraviolet by allowing one to control the lattice spacing. This is
a feature that is missing in the original lattice model where no obvious coupling
sets the lattice spacing. We investigate this new action in the next section.

\section{\label{sec7}Adding a Yang-Mills term and a TRG analysis}

In this section we will study the structure of the gravity model
in the presence of an additional Yang-Mills term.  By tuning $\beta$ larger, this pushes the plaquette group elements to be close to the identity, effectively reducing the lattice spacing.
The lattice action takes the form
\begin{align}
\label{eq:grav+ym}
    S = -\kappa \sum_{x} \Tr[M U_{x,12}] - \beta \sum_{x} \Tr[U_{x, 12}].
\end{align}
where as before $M=i\sigma_3=e^{i\sigma_3\frac{\pi}{2}}$
From our earlier Fisher zero analysis we expect any critical points to occur for
pure imaginary values of $\kappa$.  The two terms can be combined by a renormalization of $\beta$ and $\kappa$.  We can define an $SU(2)$ matrix $\tilde{M} = (\mathbf{1}\beta + i \sigma_{3} \kappa) / \mu$ with $\mu = \sqrt{\beta^2 + \kappa^2}$.  With this matrix we can re-write the action exactly as
\begin{align}
    S = -\mu \sum_{x} \Tr[\tilde{M} U_{x,12}].
\end{align}
With the action written in this form we can follow the same steps as before for the character expansion with the replacement of $\kappa$ from Sec.~\ref{sec5} by $\mu$, and $M$ with $\tilde{M}$.  This gives a nontrivial angle for the the rotation matrix, and the character $\chi^{j}(\tilde{M})$.  The angle $\theta$ is now $\theta = \arccos{\beta / \mu}$.

The tensor for this action can be written in a similar way to before with
\begin{align}
    f_{j}(\mu) = \frac{2 I_{2j+1}(\mu)}{\mu (2j+1)} \chi^{j}(\tilde{M})
\end{align}
and
\begin{align}
    T_{i j k l}^{(x)}(\mu) = 
    \begin{cases}
    f_{r}(\mu) & \text{if } i = j = k = l = r \\
    0 & \text{otherwise.}
    \end{cases}
\end{align}
This tensor is very convenient because it is very diagonal, with the only nonzero tensor elements being if all four tensor indices are identical.  Here each index is again thought of as being associated with one of the four links bounding a plaquette, so that contracting between tensors is thought of as gluing together plaquettes and summing over the representations those two plaquettes can have.

We will analytically continue $\kappa \to i \kappa$ in the tensor based on what was observed before in the absence of the Yang-Mills term, and then perform the tensor contractions.
Of course, one cannot contract the $T$ tensor with itself indefinitely because of the exponential cost in computing resources.  To cope with that we use the higher-order tensor renormalization group (HOTRG) \cite{PhysRevB.86.045139} and make truncations during contraction.

To better understand the model we consider a shifted version of the free energy, $f = \ln(\tilde{Z})$ where we have removed a regular part of the partition function corresponding to $j=0$ which doesn't affect the singular behavior close to a 
phase transition. We have also looked at the $\kappa$-derivative of
the free energy which is proportional to the mean area since $F_{12}=R_{12}-{\rm det}\,(e)$
\begin{align}
    \langle \int {\rm det}\,( e )\rangle = -\frac{\partial \ln \tilde{Z}}{\partial \kappa}.
\end{align}

\begin{figure}
    \centering
    \includegraphics[width=8.6cm]{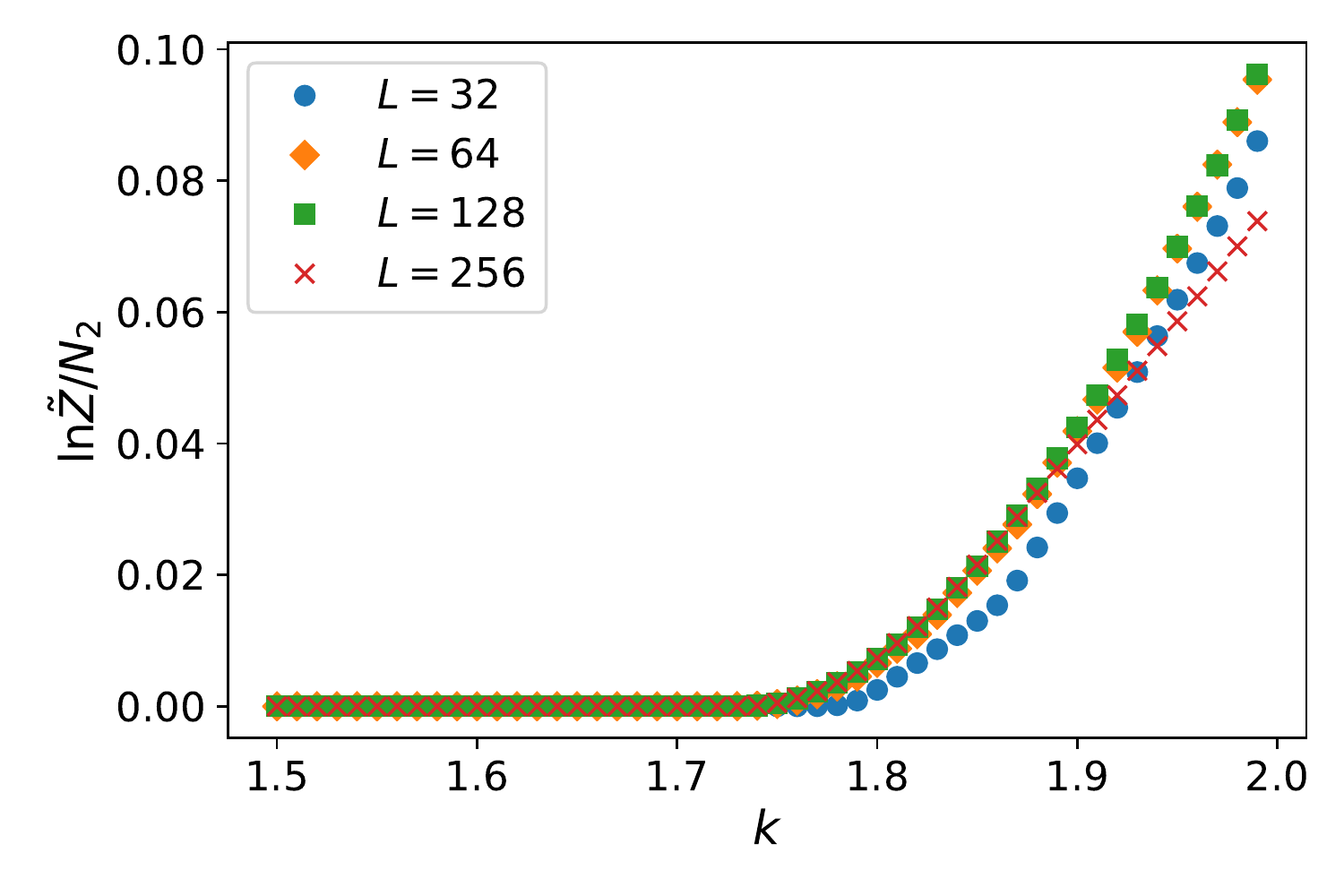}
    \caption{The normalized free energy density as a function of $k$.  Here we kept a $j_{max} = 10$, and a bond dimension of 40.  $L$ is the linear size of the lattice.}
    \label{fig:grav-ym-frengy}
\end{figure}

\begin{figure}
    \centering
    \includegraphics[width=8.6cm]{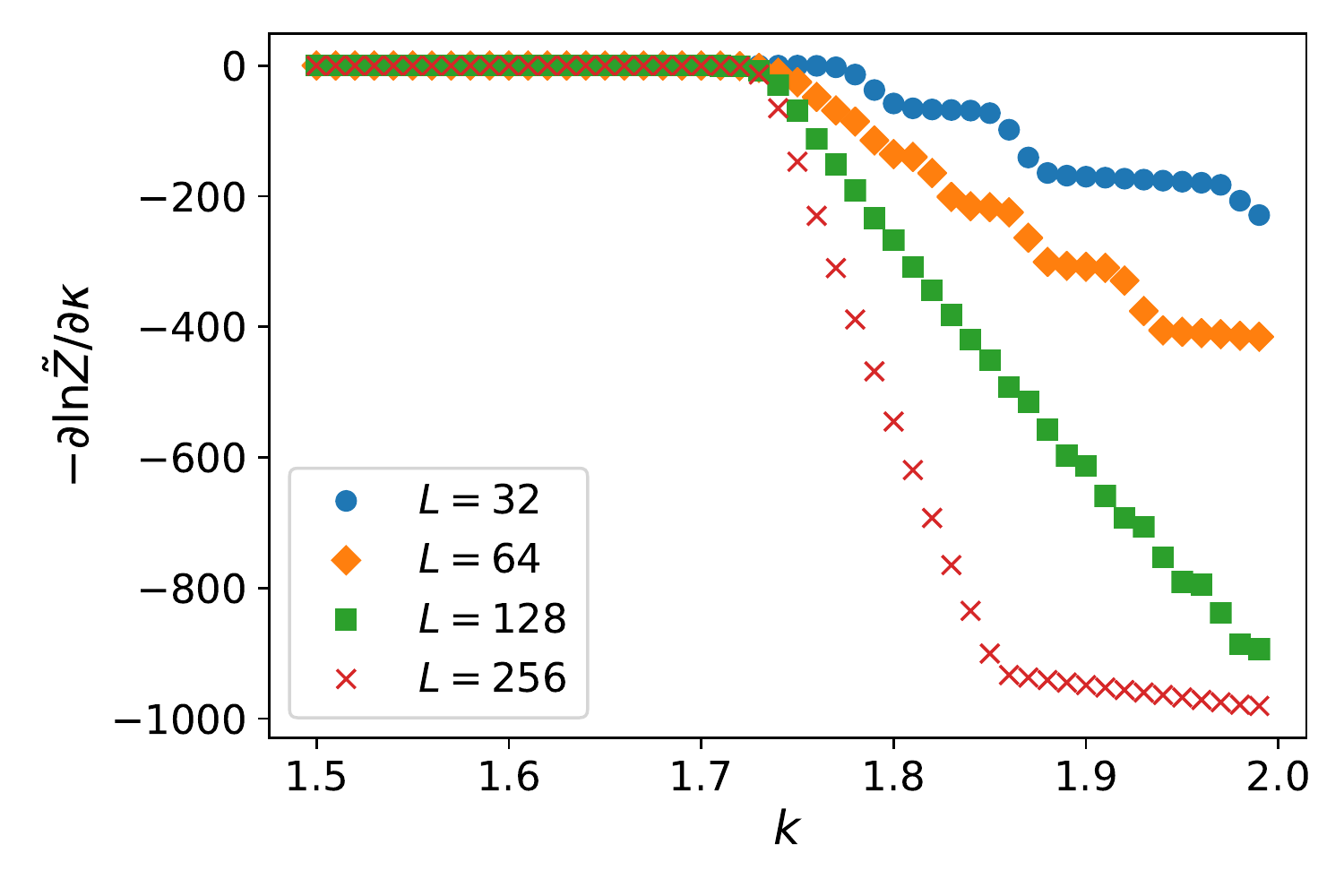}
    \caption{The quantity $\langle \int {\rm det} (e) \rangle$ as a function of the parameter $k$.  There is an abrupt change around $k \simeq 1.75$ which steepens as the volume is increased.  Here $j_{\max} = 10$, and a bond dimension of 41 states was kept.  The linear length of the lattice is $L$.}
    \label{fig:ym-grav-dete}
\end{figure}

The continuum limit is when the lattice volume is taken large while the physical volume is held fixed corresponding to keeping $\beta / N_{2} = 1/b^{2}$ with $b$ a constant, and $N_{2}$ the number of plaquettes.  We found a value of $b = 10$ was a  good value to keep stability in the size of the numbers and still see movement towards the continuum.  We also tried scaling $\kappa$ in such a way to obtain critical behavior.  We find that scaling $\kappa = k \sqrt{\beta}$, with $k$ a constant shows a transition in $k$.
The free energy density can be seen in Fig.~\ref{fig:grav-ym-frengy}.
From this figure, there appears to be a transition around $k \simeq 1.75$.  The average action can be seen in Fig.~\ref{fig:ym-grav-dete} and shows that the average area diverges for
$k \simeq 1.75$. 
In the next section, we use analytic methods to study this transition in more
detail and show how the
critical behavior depends on the topology of the lattice.

\section{\label{TRG} Analytic continuation and topology dependence}

In the previous section we demonstrated how to write a tensor network for the
lattice gravity action supplemented by a Yang-Mills term.
We showed how the two terms, the Yang-Mills and gravity terms, could be combined into a single term, then allowing for a character expansion in the $SU(2)$ characters.  We found evidence of a first-order transition in the derivative of the free energy, and here we give an analytical calculation which agrees with the numerical evidence. 
Furthermore, we are able to show the relevant critical exponent
governing the behavior of the partition function at this critical point is determined
only by the Euler characteristic of the underlying lattice. 
This feature is shared with continuum treatments of two-dimensional quantum gravity.

We start our analysis with a square lattice with torus topology, then consider a lattice
drawn on the surface of a cube with spherical topology and consider the general
case at the end, constructing a general tensor network formulation for an arbitrary random
triangulation.

\subsection{Square lattice with torus topology}
\label{subsec:torus}
Taking the action from section~\ref{sec7} we combine the two terms.
The resulting action is, again,
\beq
S = -\mu \sum_{x} {\rm Re} \Tr[\tilde{M} U_{12}(x)].
\eeq
Performing a character expansion as before and a subsequent link integration yields
\begin{align}
\nonumber
\label{eq:sqymz}
Z = &\sum_j \left[\frac{2 I_{2j+1}(\mu)}{\mu}\right]^{N_2}
(2j+1)^{N_2-N_1} \times \\
& \prod_{\rm sites\, i}\chi^j(e^{i\theta\sigma_3})
\end{align}
where, as before, the character $\chi^j$ of an $SU(2)$ representation is given by
\beq
\chi^j=\frac{\sin{\left((2j+1)\theta \right)}}{\sin{\left(\theta \right)}}.
\eeq
In the continuum it is convenient to describe the theory in terms of a dimensionless coupling $b=g L $ where $L$ is the linear size
of the system and $g$ the (dimensionful) two-dimensional Yang-Mills coupling.
In the lattice theory this
implies the continuum limit is approached by sending the number of sites along a side  of the lattice $L\to\infty$ and the lattice coupling
$g \to 0$ while holding $b=\frac{L}{\sqrt{\beta}}=g L$ fixed. This limit corresponds to taking
$\beta=\frac{N_2}{b^2}$. In the model discussed here one also needs to decide how to scale the
coupling $\kappa$. One simple choice is to hold it fixed. In this case the Yang-Mills term dominates in the continuum limit. 
This is the
situation explored in \cite{witten1991}.
In this limit $\theta\to 0$, $\chi^j=(2j+1)$, $Z$ only depends on $\beta$ and the Euler character $\chi=N_2-N_1+N_0$ and the model exhibits no phase transitions at finite coupling.
However a more interesting limit consists of scaling $\kappa\to k \frac{\sqrt{N_2}}{b}$, with $k$ a constant. In this case 
$\theta \simeq \frac{k b}{\sqrt{N_{2}}}$.
Using the small $\theta$ expansion of $\chi$ 
\beq
\chi^j\sim (2j+1)\left(1-\frac{2}{3}\theta^2 j(j+1) + \mathcal{O}(\theta^4) \right).
\eeq
we find
\beq
\chi^j \simeq (2j+1)\left(1-\frac{2}{3}j(j+1) \frac{k^2 b^2}{N_2}\right)\eeq
To construct the (singular part of the) partition function we will also normalize 
each modified Bessel function by dividing it by $I_1(\mu)/\mu$ and make use of the asymptotic
formula
\beq
\frac{I_{2j+1}(x)}{I_1(x)}=e^{-\frac{2}{x}j(j+1)},\quad x\to\infty\eeq
The final expression for the partition function as $N_2\to\infty$  is then
\begin{align}
\nonumber
Z &= \sum_j \left[e^{-\frac{2b^2}{N_2}j(j+1)}\right]^{N_2}\left(1-\frac{2 k^2 b^2}{3}j(j+1)\frac{1}{N_0}\right)^{N_0} \\
&\times (2j+1)^\chi\\
&=\sum_j (2j+1)^\chi e^{-2b^2j(j+1)\left(1+\frac{1}{3}k^2\right)}.
\label{finalZ}
\end{align}
Now consider once continuing $\kappa$ to imaginary values. This corresponds to the choice for the coupling
made in section~\ref{sec7}. In this case the expression for
$Z$ indicates a two phase structure for the system with $Z$ finite for
$k<k_c$ and divergent for $k>k_c$ where $k_c=\sqrt{3} \approx 1.73$. Furthermore, for $k<k_c$
the sum can be 
approximated  by an integral yielding a singular behavior as $k\to k_c$ 
\beq
Z\sim \left(k-k_c\right)^{-\frac{1}{2}\left(\chi+1\right)}\eeq
We thus see that the partition function has a topological character with a critical exponent depending only on the Euler number.
\footnote{It is
also interesting to notice that the form of the partition function function resembles a random surface model 
$\sum_A e^{\left(k-k_c\right)A}$ if we identify the quantized area $A=j(j+1)$ in a manner similar to loop quantum gravity.}
Of course the calculation that we have just described 
was performed on a torus where $\chi=0$. However, once might suspect that a similar result holds on more general
topologies since they can always be constructed from a square lattice using a finite number of defects. We show this explicitly in the next
section by repeating the analysis for the surface of a cube.

\subsection{Square lattice with spherical topology}
\begin{figure}
    \centering
    \includegraphics[width=0.49\textwidth]{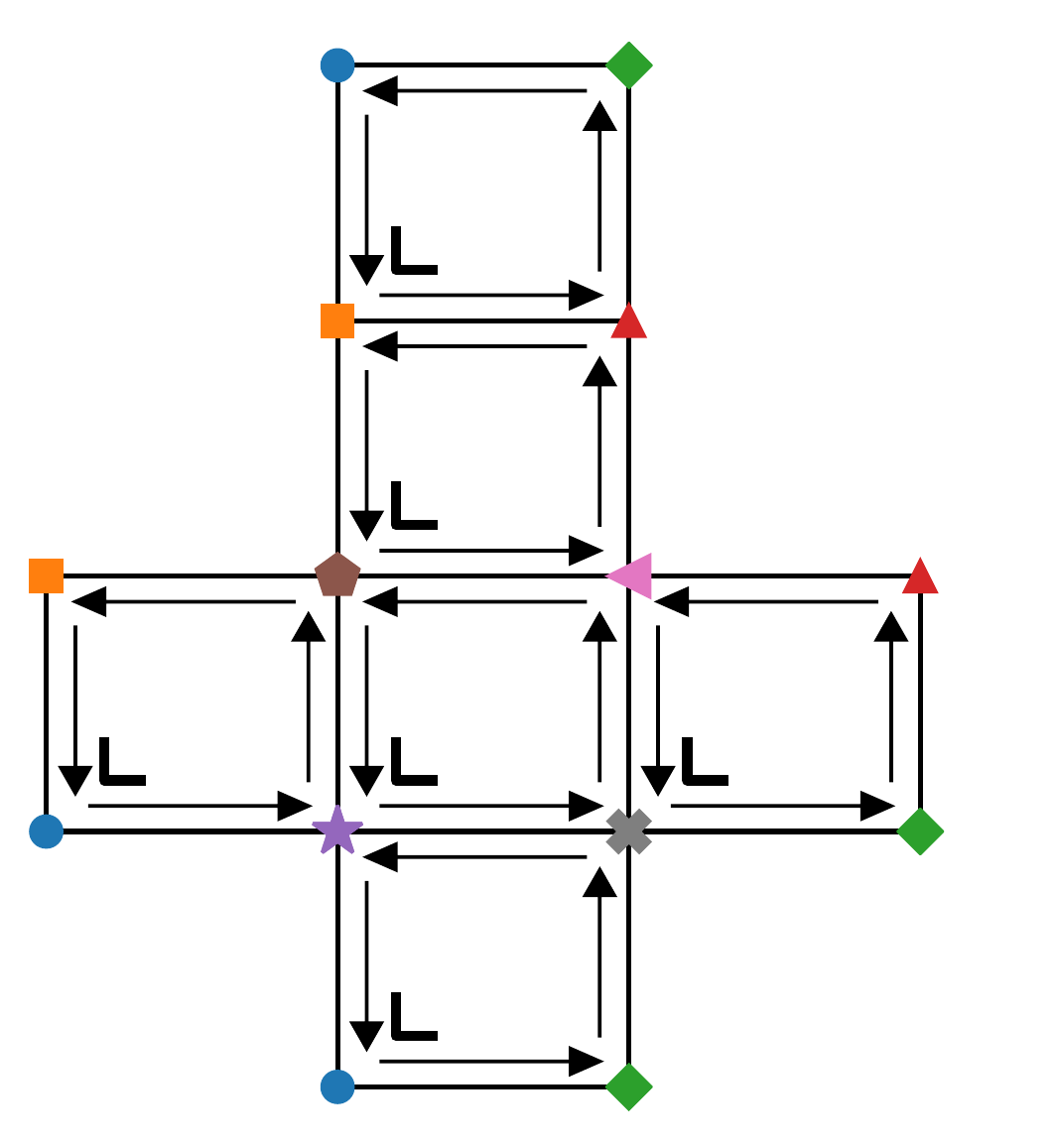}
    \caption{A figure of a $1\times 1$ cube ($L = 1$) with the orientation in mind for this article.  The same vertices are indicated with identical markers and colors, and the location of the $M$ matrix insertions are shown as well as the black inserts in the lower left-hand of the plaquette.}
    \label{fig:cube}
\end{figure}

Here we consider a lattice corresponding to the surface of a large cube.  This is a two-dimensional surface with the topology of a sphere.  The single plaquette action is the same as in Sec.~\ref{subsec:torus}; however, we must be sure to orient the surface consistently with a right-hand rule, for instance.  Fig.~\ref{fig:cube} gives an example of small cube lattice which has the same general orientation we consider in this case.

Using Fig.~\ref{fig:cube} as a guide, there are some vertices in this convention which have two $M$ matrices associated with them which are multiplied and traced over.  There are also some with no $M$ matrices associated with them; these are traces of Kronecker deltas.  And as in the torus case, there are some vertices with a single $M$ associated with them.  For a cube of edge-length, $L$, there are $L+1$ vertices on each of the 12 edges.  We find the total volume of the surface in terms of the number of vertices to be $N_{0} = 6L^{2} + 2$.  The number of vertices along two edges which share a vertex is $2L+1$.  Based on Fig.~\ref{fig:cube}, we see that there are two connected edges which contain vertices that have products of two $M$ matrices.  The two vertices at the ends of these edges however only have a single $M$ associated with them.  In addition, there are two connected edges whose vertices possess no $M$ matrices around them.  With this information we can write down the partition function for this system,
\begin{align}
\nonumber
\label{eq:cubeymz}
Z = &\sum_j \left[\frac{2 I_{2j+1}(\mu)}{\mu}\right]^{N_2}
(2j+1)^{N_2-N_1} \times \\
& \chi^j(e^{i\theta\sigma_3})^{6L^2 - 4L + 2} \chi^{j}(e^{i \theta \sigma_3 2})^{2L-1} (2j+1)^{2L+1}.
\end{align}
From this we can see that taking the same limiting behavior as before for $\kappa$ and $\beta$, we find a factor at leading order which is just
$(2j+1)^{N_{0}}$.  In addition, it turns out that the transition value of $k_c$ is unchanged on this topology (although the exponent certainly changes) which can be seen from the next leading order contributions and
we arrive back at Eq.~\eqref{finalZ}

\subsection{Arbitrary triangulation}

It is possible to write down a lattice action which can be put on 
a triangulation of any orientable closed surface and generalizes the torus and sphere constructions.  The smallest Wilson loops are
 now defined on elementary triangles where the single triangle action has the form
\begin{align}
    s_{\triangle} = - \mu \text{Re}\Tr[M^{\frac{1}{3}} U_{1} M^{\frac{1}{3}} U_{2} M^{\frac{1}{3}} U_{3}].
\end{align}
Notice that we associate the factor $M^{\frac{1}{3}}$ with each vertex of a triangle. This amounts to the replacement $\theta \to \frac{1}{3} \theta$.
The total action is just the sum of the single triangle actions for every triangle,
\begin{align}
    S = \sum_{\text{triangles}} s_{\triangle}.
\end{align}
Performing the character expansion as before and doing the group integration link by link we obtain the normalized partition function,
\begin{align}
    \tilde{Z} = &\sum_j \left[\frac{I_{2j+1}(\beta)}{I_1(\beta)}\right]^{N_2}
(2j+1)^{N_2-N_1} \times \\
& \prod_{\rm sites\, i}\chi^j(e^{i \theta \sigma_3 \frac{q_{i}}{3}}),
\end{align}
where $q_{i}$ is the number of triangles around vertex $i$.
Clearly the leading term in the expansion of the character term is independent of $q_i$ and returns the same
topological factor $(2j+1)^{N_0}$. The next to leading contribution is given by
\beq
\prod_{\rm sites, i}\left(1-\frac{2 b^2 k^2}{9 N_0}j(j+1)  q_i^2\right)\eeq
This can be exponentiated to 
\beq
e^{-\frac{2 k^2 b^2}{9}j(j+1)\langle q^ 2\rangle}\eeq 
where $\langle q^2 \rangle$ is the average squared coordination number of the triangulation. Thus the main effect of reformulating the model on
an arbitrary graph is to shift the critical coupling $k_c$. This argument confirms the existence of a phase transition in the gravity
model with topology dependent critical exponents.

\section{\label{POLY} The connection to Liouville/Polyakov gravity}
The addition of the Yang-Mills term was motivated by the use of a lattice formulation in which explicit coordinate invariance is lost.
While we have seen that this operator plays a role in generating a phase transition in the model 
it does not lead to a continuum limit in agreement with Liouville gravity \cite{polyakov103quantum}. But the list of gauge invariant operators that
are allowed in the lattice theory contains additional possibilities. One of the most obvious is a kinetic term for the
scalar field $\int {\rm Tr}\,(D\phi)^2$. Indeed, since this operator is marginally relevant in the renormalization group sense
it may indeed be important in the long distance effective theory. Of course the existence of this operator necessarily implies that
we are allowing for local fluctuations in the scalar field and so the complete lattice action becomes
\beq
S=\sum_x \kappa{\rm Re}{\rm Tr}\,\left( \phi U_{12} \right) + {\rm Tr} \, \left( D_\mu \phi \right)^2-\beta{\rm Re}{\rm Tr}\,U_{12}
\eeq
where the covariant difference operator appearing in this kinetic term is given by
\beq
D_\mu\phi(x)=U_\mu(x)\phi(x+\muhat)U^\dagger_\mu(x)-\phi(x)\eeq
Notice we have used the freedom to rescale $\phi$ to set the coefficient of the derivative term to unity.
It is instructive to again examine this lattice action in unitary gauge where the action takes the form
\beq
S=\sum_x \frac{\kappa}{\sqrt{2}}\rho{\rm Re}\Tr\left(i\sigma_3U_{12}\right) +
\rho(-\Box_c)\rho-\beta{\rm Re}{\rm Tr}\,U_{12}\label{lattice}\eeq
where the Laplacian operator acts on the field $\rho$ as
\beq
\Box_c\rho(x)=\sum_\mu V_\mu(x)\rho(x+\mu)+V_\mu(x-\mu)\rho(x-\mu)-2\rho(x)\eeq
and
\beq
V_\mu(x)=\frac{1}{2}\Tr\left(\sigma_3U_\mu(x)\sigma_3U^\dagger_\mu(x)\right)\eeq
Setting $U_\mu(x)=e^{i\sum_{i=1}^3A_\mu^i(x)\sigma^i}$ we find for small $A_\mu^i$ ($\beta\to\infty$)
\beq
V_\mu(x)=1-\left(A_\mu^a\right)^2+\ldots\eeq
where the index $i\equiv a$ now runs only over the first two components of the field
since $A_\mu^3$ remains massless in this gauge.
Thus, in unitary gauge, the naive continuum limit of this
new action reads 
\beq
S=\int d^2x \left[ \kappa\sqrt{2}\rho F_{12}^3+
\rho\left(-\overline{\Box}+m^2\right)\rho+\frac{1}{g^2}\Tr(F_{12}F_{12})\right]\label{higgs}\eeq
with $\overline{\Box}$ the usual free Laplacian in flat space, 
$m^2=2\sum_\mu \left(A_\mu^a\right)^2$ and $g^2$ the
dimensionful gauge coupling in two dimensions. Holding the dimensionless parameter $b=g L$ fixed as the size $L \to \infty$ 
drives $g$ and hence the magnitude of the gauge fields to zero. In this limit then $m^2\to 0$ and $F_{12}^3\to \frac{b}{L}R_{12}^3$.
If we simultaneously scale $\kappa=k \frac{L}{b}$ as before and use the relation $\int R_{12}=\int\sqrt{g}R$
the naive continuum limit of the scalar field dependent part of the lattice
action is
\beq
S=\int d^2x\, k\sqrt{g}R\rho+\rho(-\overline{\Box})\rho.
\eeq
This takes the form of the celebrated Liouville action of two-dimensional quantum gravity if we identify the field $\rho$ with
the conformal mode and then use the result $\overline{\Box}=\sqrt{g}\Box$. Of course we expect that the action is only conformal for a particular value of
$k$ determined by the conformal anomaly. This should translate into the statement that the lattice model should possess a phase
transition at a critical value of $k_c$. Monte Carlo and TRG studies are currently underway to verify this conjecture \cite{us}.
Notice that while the connection between the lattice gauge theory and Liouville gravity has been made in unitary
gauge the correspondence will also hold in the original model containing the full scalar field $\phi$. Indeed the effective action
for the gauge fields having integrated out the scalar field $\phi$ takes the non-local Polyakov form
\beq
\frac{1}{4k^2}\int d^2x d^2y\, F_{12}(x)\Box^{-1}(x-y) F_{12}(y).
\eeq

\section{\label{sec8}Conclusions}
In this paper we have shown how to construct a lattice gauge theory whose
naive continuum limit yields a path integral for Cartan-Palatini gravity.
In two dimensions we show using character expansions
that the lattice model can be rewritten in terms of a discrete set of
dual variables and recast as a tensor network. With gravity rewritten in this language
it is possible to compute observables using tensor renormalization group 
techniques. This is useful since we have shown that typical critical points in these
gauge theoretic approaches to gravity are to
be found only for values of the gravitational coupling where the Euclidean action develops a
sign problem eliminating the possibility of using Monte Carlo simulation.

In two dimensions additional tricks based
on analytic continuation allow one to bypass the TRG techniques and find exact
solutions in certain cases. Ultimately this is likely tied to the topological nature of two
dimensional gauge theories. Indeed, we find critical exponents that depend
only on the topology of the underlying lattice.
However these exact methods will fail in higher dimensions and TRG methods
will then be critical.
For example, in four dimensions, the 
discrete analog of Eq.~\eqref{unitary} is given by \cite{Catterall:2009nz}
\beq
S=\kappa\sum_x \epsilon_{\mu\nu\rho\lambda}{\rm Tr\,} \left(\gamma_5 \left[U^P_{\mu\nu}-U^{P\dagger}_{\mu\nu}\right]\left[U^P_{\rho\lambda}-U^{P \dagger}_{\rho\lambda}\right]\right)\eeq
where the link fields are now valued in $Spin(5)$ and unitary gauge has again
been employed. The character expansion method can still be used and the
gauge fields integrated out leading to a theory
written in discrete dual variables corresponding to 
Clebsch-Gordon coefficients of $Spin(5)$. In principle this dual theory can be realized
as a tensor network and solved using TRG techniques. Of course to realize this
possibility will require substantial improvements in current TRG algorithms which
currently cannot
reach four dimensions because of memory limitations but there is no problem of principle and we
hope to report on such work in the near future.

More subtle is the question of how to
change the sign of the cosmological constant. Even in two-dimensional Euclidean
space an attempt to study anti de Sitter space would necessitate replacing the compact group $SU(2)$ by
its non compact cousin $SU(1,1)$. The latter possesses unitary representations
labeled by a continuous index in addition to a discrete series of representations which are analogs of those in
$SU(2)$. This renders the character expansion and subsequent Haar integration a much more subtle enterprise.

\acknowledgments This work is supported in part by the U.S.~Department of Energy, Office of Science, Office of High Energy Physics, under Award Number DE-SC0019139. The authors are grateful for discussions with Jay Hubisz
and Raghav Jha.

%

\end{document}